\documentclass[letter,10pt]{IEEEtran}

\makeatletter
\let\IEEEproof\proof
\let\IEEEendproof\endproof
\let\proof\@undefined
\let\endproof\@undefined
\makeatother

\usepackage{amsmath, amssymb, amsfonts, amsthm}
\usepackage{eso-pic, graphics, graphicx, xspace, epic, eepic}
\usepackage{color}
\usepackage{makeidx}

\newtheorem*{theorem*}{Theorem}
\newtheorem*{proposition*}{Proposition}
\newtheorem*{corollary*}{Corollary}
\newtheorem*{lemma*}{Lemma}
\newtheorem*{property*}{Property}

\theoremstyle{definition}

\newtheorem*{exercise*}{Exercise}
\newtheorem*{example*}{Example}
\newtheorem*{definition*}{Definition}

\theoremstyle{remark}

\newtheorem*{note*}{Note}

\let\proof\IEEEproof
\let\endproof\IEEEendproof

\usepackage[nocompress]{cite}

\begin{document}

\title{\Large\bfseries Statistical Intercell Interference Modeling for Capacity-Coverage Tradeoff\\ Analysis in Downlink Cellular Networks}

\author{
  \authorblockN{Naeem Akl, Jihad Fahs, Zaher Dawy}\\
  \authorblockA{Electrical and Computer Engineering, American University of Beirut, Beirut, Lebanon\\
   Emails: {\{nka16, jjf03, zaher.dawy\}@aub.edu.lb}
   }
    }

\maketitle

\begin{abstract}
Interference shapes the interplay between capacity and coverage in cellular networks. However, interference is non-deterministic and depends on various system and channel parameters including user scheduling, frequency reuse, and fading variations. We present an analytical approach for modeling the distribution of intercell interference in the downlink of cellular networks as a function of generic fading channel models and various scheduling schemes. We demonstrate the usefulness of the derived expressions in calculating location-based and average-based data rates in addition to capturing practical tradeoffs between cell capacity and coverage in downlink cellular networks.
\end{abstract}

\IEEEpeerreviewmaketitle

\section{Introduction}\label{Introduction}

Intercell interference (ICI) is a key factor in the design and implementation of wireless cellular networks. Deriving closed-form expressions for ICI in various wireless network scenarios has gained notable attention in the literature, as it facilitates in depth performance analysis, aids designers in developing and evaluating advanced enhancement techniques, and reduces the need for time consuming Monte-Carlo simulations.

The main contribution of this work is the analytical modeling of the statistics of downlink ICI and data rates as a function of general {\it channel models} and different {\it user scheduling schemes}; the derived models are used to capture cell {\it capacity-coverage tradeoffs}. The authors in~\cite{SC09},~\cite{SCH10},~\cite{PZG11} focused on the distribution of downlink ICI as a function of different channel models including pathloss, Rayleigh/Ricean fading, with/without log-normal shadowing. Other works presented a low-complexity approximated approach in~\cite{CIN07} and an efficient method assuming non-fading channels in~\cite{AFF11}. In~\cite{Andrews11}, the authors provided general models for the multi-cell signal-to-interference noise ratio~(SINR) using an approach based on stochastic geometry. Assuming that the base stations~(BSs) positions are independent and distributed according to a Poisson point process, they qualify their model as being pessimistic and equally accurate as deterministic models; furthermore, they utilized their approach to present capacity-coverage tradeoff results without taking into account the effect of scheduling. In this work, we consider a more deterministic system model with dependence among BS locations and we capture analytically the impact of user scheduling on cell capacity-coverage performance. In~\cite{TYD13}, ICI models are derived for different types of scheduling schemes using a semi-analytical approach, however, considering an uplink scenario. To the authors' best knowledge, no previous work has presented analytical statistical models to capture  capacity-coverage tradeoffs in downlink wireless networks as a function of user scheduling for general fading~channels.

The rest of this paper is organized as follows: In Section II, we present the general system model. In Section III, we consider a single cell channel model and derive the corresponding downlink cell rate distribution under three types of scheduling schemes, namely the round-robin, greedy and proportional-fair. Then, we extend the derivations in Section IV to capture ICI in a multi-cellular network scenario. Finally, we present in Section V closed form expressions and capacity-coverage tradeoff results for a case study scenario under various scheduling schemes and Section VI concludes the paper.

\section{System Model}
\label{SM}
We consider a single carrier in an OFDMA-based downlink network scenario. Cells are modeled as circles of radius $\rho$ where the BSs are located at the cell centers. We assume that the central cell is surrounded by multiple BSs that act as sources of interference; without loss of generality, we restrict the number of interferers to six considering only the first tier of cells, and assume that BSs are equidistant from the serving BS with a distance of $2\rho$. $N$ users are present in each cell, and distributed independently of the angle $\theta$, according to $f_D(\delta) = \frac{2\delta}{\rho^2}$, where $\delta$ is the distance between a given user and the serving BS. In the general setting, the serving BS is assumed to transmit with a given power $P$ and we denote by $P_j$ the power allocated at the interfering BS of cell $j$, $1 \leq j \leq 6$. Transmitted signals are subjected to AWGN, distance-based pathloss, and fading. Pathloss is modeled as $10^{\text{K}/10}(\frac{\delta}{\text{d}_0})^{-\alpha} = \xi \delta^{-\alpha}$, where $d_0$ is reference distance, $\alpha$ is the pathloss exponent and $K$ in dB is the pathloss constant. We couple the serving BS transmit power together with the pathloss constant, reference distance and the noise power into the parameter $\xi$. Similarly, the interfering signals suffers from a pathloss of the form $\xi_j \delta_j^{-\alpha}$ where  $\xi_j = 10^{\frac{K_j}{10}} \, P_j \, d_0^{\alpha_j}$ and $\delta_j$ being the distance between the BS of cell $j$ and the user. $\alpha_j$ and $K_j$ are the pathloss exponent and pathloss constant for the channel between cell $j$ and the user. For a user located in the central cell, a generic fading model of the signal of interest is considered that can be represented via a composite fading distribution that captures both fast fading and shadowing; it is modeled via the random variable $A$ with probability density function~(PDF) $f_A(a)$. The interfering channels are also assumed to be subject to independent fading statistics $A_j$, $1 \leq j \leq 6$. We assume that the fading statistics are identically independently distributed (iid) among users.

\begin{figure}[t!]
  \begin{center}
    \includegraphics[width=3.4in]{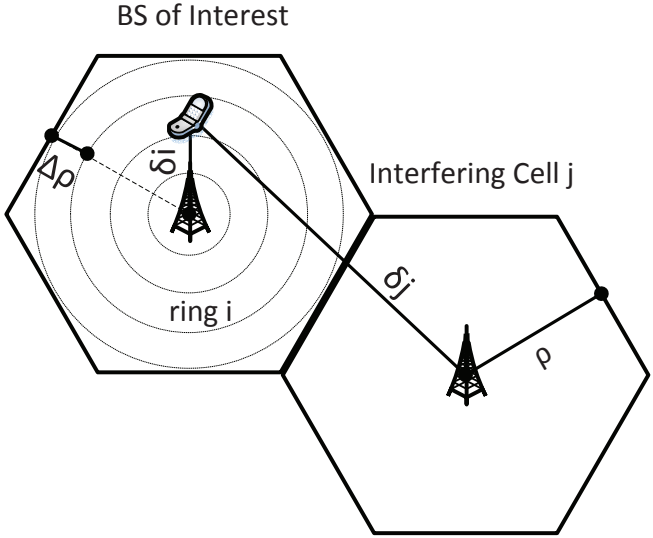}
    \caption{\small General channel model with one interfering base station
      \label{figure0}}
  \end{center}\vspace{-0.4cm}
\end{figure} 

\section{Single Cell Scenario: Rate Distribution}
\label{TA}
The single-cell scenario is a special case of the general model where no interfering BSs exist. The signal-to-noise ratio~(SNR) for a $\delta$-distant user from the BS can be expressed as $\Gamma = \xi \,\delta^{-\alpha}\, A $. In this section, we derive the PDF of the downlink rate considering three types of scheduling schemes, namely, round-robin, proportional fair and greedy.

\subsubsection{Round-Robin}
Round-robin scheduling allocates all resources to a cell user chosen arbitrarily and, thus, provides an extreme case focused on fairness. Hence the selection process of the served users has the same statistics as $f_D(\delta)$. For a selected user at distance $\delta$ from the BS, the downlink rate is given by Shannon's capacity relation for AWGN channels~\cite{Sha48_1,Sha48_2}:
\begin{equation}
 R = \log (1 + \xi\,\delta^{-\alpha} \, A)
\label{cap}
\end{equation}
which gives
\begin{equation}
f_{R/D}(r / \delta) = \frac{\delta^{\alpha}}{\xi}e^{r} f_{A}\left(\frac{(e^r-1)\delta^{\alpha}}{\xi}\right).
\end{equation}
Averaging over the user distribution in the cell would yield the following downlink rate PDF under round-robin scheduling:
\begin{equation}
f_{R} (r) = \int_{0}^{\rho} f_{D}(\delta) f_{R/D} (r /\delta) \, d\delta
\label{pdfRR}
\end{equation}

\subsubsection{Greedy}
Greedy scheduling allocates resources to the best user and, thus, provides the other extreme case focused on throughput maximization. Thus, the user with the maximum SNR is served and~(\ref{cap}) becomes:
\begin{equation}
R = \log(1 + \Gamma_{\max}).
\label{grcap}
\end{equation}
where the distribution of the maximum SNR $\Gamma_{\max}$ is to be determined.
Using a similar approach as the one proposed in~\cite{TYD13}, we start by discretizing the cell into $M$ rings. Then, we proceed by finding the best user within a given ring. Let $\Gamma_{\max, i}$ be the maximum SNR within ring $i$, $1 \leq i \leq M$ which corresponds to the maximum fading $A_{\max, i}$ having the following cumulative distribution function~(CDF):
\begin{equation}
F_{A_{\max, i}}(a)= \prod_{j=1}^{N_{i}} F_{A_j}(a) = F_{A}^{N_{i}}(a),
\label{maxa}
\end{equation}
where $N_i$ is the number of users in ring $i$. One should note that $\{N_{i}\}_{1 \leq i \leq M}$ is a binomial RV $\mathcal{B}(n,p)$ with parameters $n = N$ and the ratio of the areas of the ring and the cell $p = \frac{2 \delta_i \Delta \rho}{\rho^2}$. $\Delta \rho$ is the difference between the small and the large radii of the rings and $\delta_i$ is the distance between the center of ring $i$ and the cell center(see figure~\ref{figure0}). It is known that the mean value of a RV $X \sim \mathcal{B}(n,p)$ is $np$ and its variance is $np(1-p)$. This would imply that the variance of $N_i$ can be made arbitrarily small by the choice of $\Delta \rho$ and therefore one could approximate $N_i$ by its mean value $\text{E}[N_i]  = N \frac{2 \delta_i \Delta \rho}{\rho^2}$ which is assumed to have integer values for $N$ large enough. Hence, one can evaluate the distribution of the maximum SNR as follows:
\begin{equation}
\label{gammax}
F_{\Gamma_{\max}}(\gamma)= \prod_{i=1}^{M} F_{A}^{\text{E}[N_{i}]}\left(\frac{\delta_{i}^{\alpha}\gamma}{\xi} \right),
\end{equation}
 As $M \rightarrow \infty$, we get:
\begin{equation}
\log F_{\Gamma_{\max}}(\gamma) = N\,\int_0^{\rho} \frac{2\delta}{\rho^2} \log F_{A} \left(\frac{\delta^{\alpha}\gamma}{\xi} \right) \,d\delta.
\label{eqmaxsnrcellgen}
\end{equation}
Using~(\ref{grcap}), the downlink rate PDF under greedy scheduling is given by:
\begin{equation}
f_{R}(r) = e^r\frac{dF_{\Gamma_{\max}}}{d\gamma}\left(e^r-1\right).
\label{Gr_Jac}
\end{equation}

\subsubsection{Proportional-Fair}
\label{PF}
Proportional fair scheduling schemes are used to balance the tradeoff between fairness, throughput, and effective cell coverage. To find the rate distribution with a proportional fair scheme, we proceed as follows:
\begin{itemize}
\item Select the user with the maximum normalized SNR:
\begin{equation}
\label{selpro}
 A_{\text{sel}} = \max \limits_{ 1 \leq i \leq N} \left\{\frac{\Gamma_{i}}{\overline{\Gamma}_{i}}\right\} = \max \limits_{ 1 \leq i \leq N} \left\{\frac{A_{i}}{\overline{A}_{i}}\right\}
\end{equation}
where $\Gamma_i$ and $A_i$ are the SNR and the fading values for user $i$, respectively; $\overline{\Gamma}_{i}$ and $\overline{A}_{i}$ are the corresponding average values. 
\item This would reduce, under the assumption of identical fading statistics, to choosing the user having the maximum fading. Hence,
\begin{equation}
F_{A_{\text{sel}}}(a) = \prod_{i=1}^{N}F_{A_{i}}(a) = F_{A}^{N}(a),
\end{equation}
and
\begin{equation}
f_{A_{\text{sel}}}(a) = N  f_{A}(a) F_{A}^{N-1}(a).
\end{equation}
\item Thus, the rate can be expressed as
\begin{equation}
\label{pfres}
R=\log(1 + \xi\,\delta_{\text{sel}}^{-\alpha} \, A_{\text{sel}})
\end{equation}
where $\delta_{\text{sel}}$ is the distance of the selected user to the BS. Considering (\ref{selpro}), it can be seen that the selection process is uniform among all users; hence, $f_{\delta_{\text{sel}}(\delta)} = \frac{2\delta}{\rho^2}$. Finally, conditioning (\ref{pfres}) for a fixed $\delta_{\text{sel}} = \delta$ and averaging over $f_{\delta_{\text{sel}}(\delta)}$ gives the following rate distribution:
 \begin{equation}
 f_{R}(r) = \frac{2e^{r}}{\xi\rho^2}\int_{0}^{\rho} \delta^{\alpha  + 1}f_{A_{\text{sel}}}\left(\frac{\delta^{\alpha}}{\xi}(e^{r}-1)\right)\,d\delta.
 \end{equation}
\end{itemize}

\section{Multi-Cell Scenario: ICI and Rate Distributions}

At a given user location $(u,v)$ in the central cell, the total received interfering power can be expressed as:
\begin{equation}
\text{I}_{\text{total}}=\sum_{j=1}^{6}\,\xi_j\,\delta_{j}^{-\alpha_{j}}\,A _{j} = \sum_{j=1}^{6} \text{I}_{j},
\label{totint}
\end{equation}
which translates using the independence of the $A_j$'s to
\begin{equation}
\text{MGF}_{\text{total}}=\prod_{j=1}^{6}\text{MGF}_{j},
\label{MGF_total}
\end{equation}
where $\text{MGF}_{j}$ is the moment generating function of $\text{I}_j$. 
For a user located at $(u,v)$, the SINR $\Gamma$ can be expressed as follows:
\begin{equation}
\Gamma(u,v)=\frac{\xi\,\delta^{-\alpha}\,A}{\text{I}_n+\text{I}_{\text{total}}},
\end{equation}
where $\delta$ is the distance of user $(u,v)$ from the central BS, $\xi = 10^{\frac{K}{10}} \, P \, d_0^{\alpha}$ is the composite power at the central BS, $A$  is the composite fading, and $\text{I}_n$ is the additive noise power. Using the distribution of the total interfering power $f_{\text{I}_{\text{total}}}(\cdot)$ at user location $(u,v)$, its SINR PDF can be derived as:
\begin{equation}
f_{\Gamma}\left(\gamma|(u,v)\right) =\int_{0}^{\infty}\frac{n+\eta}{\xi\,\delta^{-\alpha}}f_{A}\left(\frac{(n+\eta)\gamma}{\xi\,\delta^{-\alpha}}\right)f_{\text{I}_{\text{total}}}(\eta)\,d\eta.
\end{equation}
Moreover, the rate PDF at any location $(u,v)$ inside the cell of interest can be derived as follows:
\begin{equation}
f_{R}\left(r|(u,v)\right)=\, e^{r}\,f_{\Gamma}\left(e^{r}-1|(u,v)\right).
\label{ratexy}
\end{equation}
In order to {\em analytically capture the impact of a wide range of scheduling schemes on downlink network performance}, we propose an approach based on using a truncated Gaussian distribution with variable variance to model user selection density. More precisely, we model the scheduler's user selection density inside a cell as follows:
\begin{equation} \label{eq:gen}
f_{D}(\delta) = \frac{1}{\beta}\frac{\delta}{\sigma^{2}}e^{-\frac{\delta^{2}}{2\sigma^{2}}}
\end{equation}
where $\beta=e^{\frac{-d_0^2}{2\sigma ^{2}}}-e^{\frac{-\rho^{2}}{2\sigma ^{2}}}$ is a normalizing factor. Using (\ref{eq:gen}), we can emulate a wide range of scheduling schemes including proportional fair scheduling by varying the parameter $\sigma$, where $\sigma \in [0,\infty[$. The higher the value of $\sigma$, the more uniform is the distribution and, thus, more fair the user selection; the round-robin scheme can be obtained by letting $\sigma \rightarrow \infty$. The smaller the value of $\sigma$, the higher the density of users selected towards the cell center and, thus, the higher is the cell throughput. Note that the greedy scheme does not map simply to the case $\sigma \rightarrow 0$. However, one can find a value of $\sigma$ for which the proposed distribution coincides with greedy scheduling; this is channel specific as discussed in the next section.

\section{Performance Results}
In this section, we utilize the derived distributions, which are valid for general fading models, to study analytically the cell rate and capacity-coverage tradeoffs in downlink cellular networks considering a Rayleigh channel where we were able to obtain closed form expressions. We assume the following set of parameters, without loss of generality: Rayleigh channel model with unit average power $f_A(a) = e^{-a}$, $P = 1$ W, $K = -80$~dB, $d_0 = 1$~m, $\alpha = 2$, bandwidth $W = 10$~KHz, and $\text{I}_n = 10^{-14}$~W/Hz. For a cell radius of $\rho = 1000$~m, we consider that the number of users is equal to $N=100$ users and we maintain a fixed user density as $\rho$ varies.\vspace{-0.25cm}

\subsection{Single Cell Scenario: Results and Analysis}
For the single-cell scenario, an evaluation of~(\ref{pdfRR}) assuming round-robin scheduling gives:
\begin{equation}
f_{R}(r) = \frac{e^{r}}{e^{r}-1}\left[\frac{1-e^{-\frac{\kappa}{\xi}}}{\frac{\kappa}{\xi}}-e^{-\frac{\kappa}{\xi}}\right],
\label{CLRR}
\end{equation}
where $\kappa=\rho^2(e^r-1)$. As for the greedy scheme,~(\ref{eqmaxsnrcellgen}) becomes:
\begin{equation}
\log F_{\Gamma_{\max}}(\gamma) = \frac{2N}{\rho^2}\int_0^{\rho} \delta\,\log \left(1-e^{-\delta^{2} \gamma} \right) d\delta.
\end{equation}
Moreover, (\ref{Gr_Jac}) can be evaluated for extreme low and high values of $r$ as follows:
\begin{equation}\label{CLGr}
  f_{R}(r) = \left\{ \begin{array}{ll}
      \displaystyle \frac{N}{e^N} \left(\frac{\rho^2}{\xi}\right)^N r^{N-1} \qquad \qquad \quad \,\,\,\,\,\,\,\,r \rightarrow 0\\
       \displaystyle N  \frac{\xi}{\rho^2} e^{-r} \,\,\, \,\,\, \,\qquad \qquad \qquad \qquad \quad  r \rightarrow \infty
    \end{array} \right.
\end{equation}
Comparing (\ref{CLRR}) with (\ref{CLGr}) it can be shown that the probability of having low rates under greedy scheduling is much lower than that of round-robin, whereas the probability of having higher rates is $N$ times more; this quantifies in {\em closed-form} the multiuser diversity gain achieved by greedy scheduling.

\begin{figure}[t!]
  \begin{center}
    \includegraphics[width=3.4in]{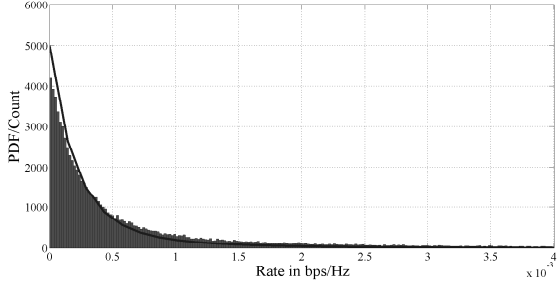}
    \caption{\small Analytical and empirical rate PDF for the round-robin scheme. The empirical PDF is presented using bar diagram.
      \label{figure1}}
  \end{center}\vspace{-0.4cm}
\end{figure}

\begin{figure}[t!]
  \begin{center}
    \includegraphics[width=3.4in]{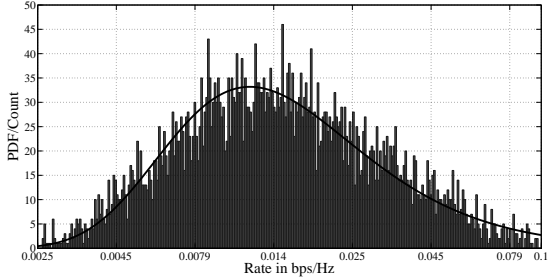}
    \caption{\small Analytical and empirical rate PDF for the greedy scheme. The empirical PDF is presented using bar diagram.
      \label{figure2}}
  \end{center}\vspace{-0.4cm}
\end{figure}

\begin{figure}[t!]
  \begin{center}
    \includegraphics[width=3.4in]{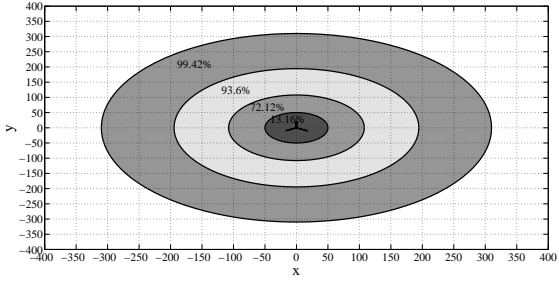}
    \caption{\small Effective coverage under greedy scheduling showing the cumulative percentage of the overall served users that are located within a given radius from the cell center.
      \label{figure3}}
  \end{center}\vspace{-0.4cm}
\end{figure}

In Fig.~\ref{figure1} and Fig.~\ref{figure2}, we plot the rate PDFs for round-robin and greedy scheduling, respectively, assuming a cell radius of $\rho = 1000 \, \text{m}$. Moreover, we plot the corresponding empirical rate distributions using Monte-Carlo simulations for verification purposes. The results quantify analytically the data rate variation as a function of the user scheduling scheme.
The enhanced rate performance of greedy scheduling comes at the expense of cell coverage. In Fig.~\ref{figure3}, we show that approximately $99.4\%$ of the selected users are within an effective radius of $300$~m from the BS and hence only $9\%$ of the cell area is effectively covered. In practice, proportional fair scheduling schemes are used to better balance the tradeoff between fairness, throughput, and effective cell coverage.

 \subsection{Multi-Cell Scenario: Results and Analysis}
\begin{figure}[t!]
  \begin{center}
    \includegraphics[width=3.4in]{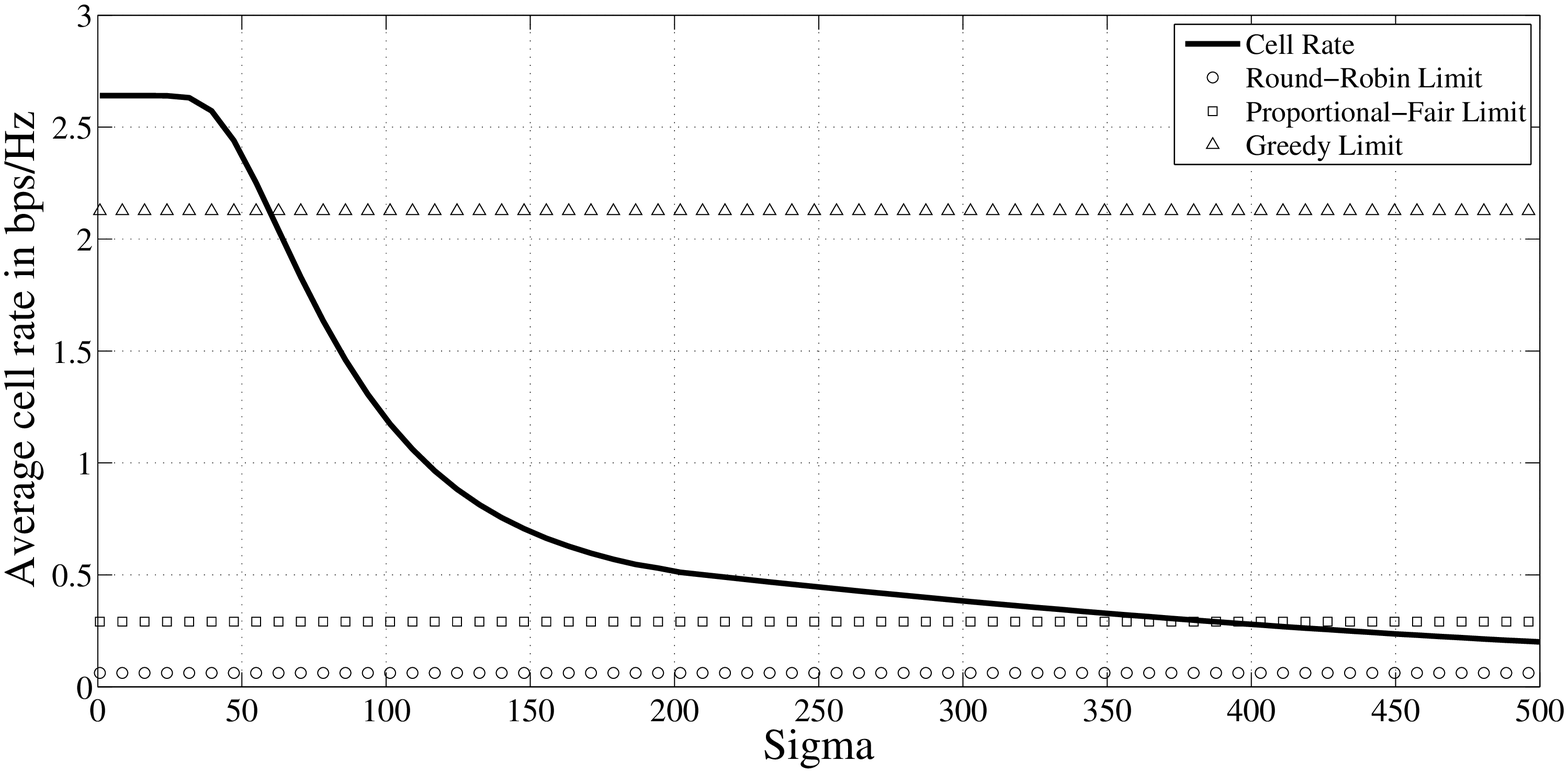}
    \caption{\small Average cell rate versus scheduler $\sigma$ parameter in a multi-cell network scenario, $\rho = 1000 \, \text{m}$.
      \label{figure4}}
  \end{center}\vspace{-0.4cm}
\end{figure}

 Assuming a Rayleigh fading channel model, (\ref{totint}) is the sum of six exponential random variables, each of average $\overline{\text{I}}_{j}=\delta_{j}^{-\alpha}\,\xi_j$. Moreover, the signal of interest at user $(u,v)$ is also exponential with average $\overline{\text{I}}=\delta^{-\alpha}\,\xi$. Equation~(\ref{MGF_total}) can be simplified as follows:
\begin{equation}
\text{MGF}_{\text{total}} = \prod_{j=1}^{6}\frac{1}{1-s\overline{\text{I}}_{j}},
\end{equation}
which gives by taking the inverse Laplace transform
\begin{equation}
f_{\text{I}_{\text{total}}}\left(\eta|\left(u,v\right)\right) =\sum_{j=1}^{6}\frac{\text{C}_{j,6}}{\overline{\text{I}}_{j}}e^{-\frac{\eta}{\,\,\overline{\text{I}}_{j}}},
\end{equation}
where
\begin{equation}
\label{coef}
C_{j,M} =\prod_{l=j+1}^{M}V_{j,l}\cdot\sum_{k=1}^{j-1}V_{j,k}C_{k,j-1},
\end{equation}
and $V_{j,l} = \frac{\overline{\text{I}}_{j}}{\overline{\text{I}}_{j} - \overline{\text{I}}_{l}}$. We note that in~(\ref{coef}), whenever the indices are such that $m>n$, the expressions $\sum_{l=m}^{n}$ and $\prod_{l=m}^{n}$ are set to~1. Hence, the downlink rate distribution at location $(u,v)$ in~(\ref{ratexy}) can be simplified as follows:
\begin{equation}
f_{R}\left(r|(u,v)\right) =e^r e^{-\frac{(e^r-1)\,\text{I}_n}{\overline{\text{I}}}}\sum_{j=1}^{6}C_{j,6}\left(c+d\right),
\label{pdf_rate_x_y}
\end{equation}
where
\begin{eqnarray*}
c &=&\frac{\text{I}_n}{(e^r-1)\overline{\text{I}}_{i}+\overline{\text{I}}},\\
d &=&\frac{x\delta^{-\alpha}\overline{\text{I}}_{i}}{\left((e^r-1)\overline{\text{I}}_{i}+\overline{\text{I}}\right)^2}.
\end{eqnarray*}
For interference-limited regimes, the average downlink rate can be derived in closed form as follows:
\begin{equation}
\overline{R}(u,v) =\sum_{j=1}^{6}\frac{\overline{\text{I}}}{\overline{\text{I}}-\overline{\text{I}}_{j}} \,C_{j,6}\,\ln\left(1+\frac{\overline{\text{I}}-\overline{\text{I}}_{j}}{\overline{\text{I}}_{j}} \right).
\label{avrate}
\end{equation}
Though in this analysis we only considered users where $\overline{\text{I}}_{i} \neq \overline{\text{I}}_{j}$, $i \neq j$, the same method can be used to derive rate distribution for the case with $\overline{\text{I}}_{i} = \overline{\text{I}}_{j}$. However, since the latter case occurs  with arbitrarily low probability, it will not affect the subsequent analysis and hence will not be considered.
Evaluating~(\ref{avrate}), Fig.~\ref{figure4} presents the average cell rate as a function of the scheduling fairness parameter $\sigma$; the average rate decreases as $\sigma$ increases and tends to the performance of round-robin scheduling as $\sigma \to \infty$. For the greedy scheme, the value of $\sigma = 58.82$ is chosen in such a way that if we nullify the interference, we obtain the same average rate and effective coverage as in the single cell scenario. The same applies for the proportional-fair scheme where a value of $\sigma = 378.67$ is considered as equivalent.

\subsubsection{Capacity-Coverage Tradeoff Results}
\begin{figure}[t!]
  \begin{center}
    \includegraphics[width=3.4in]{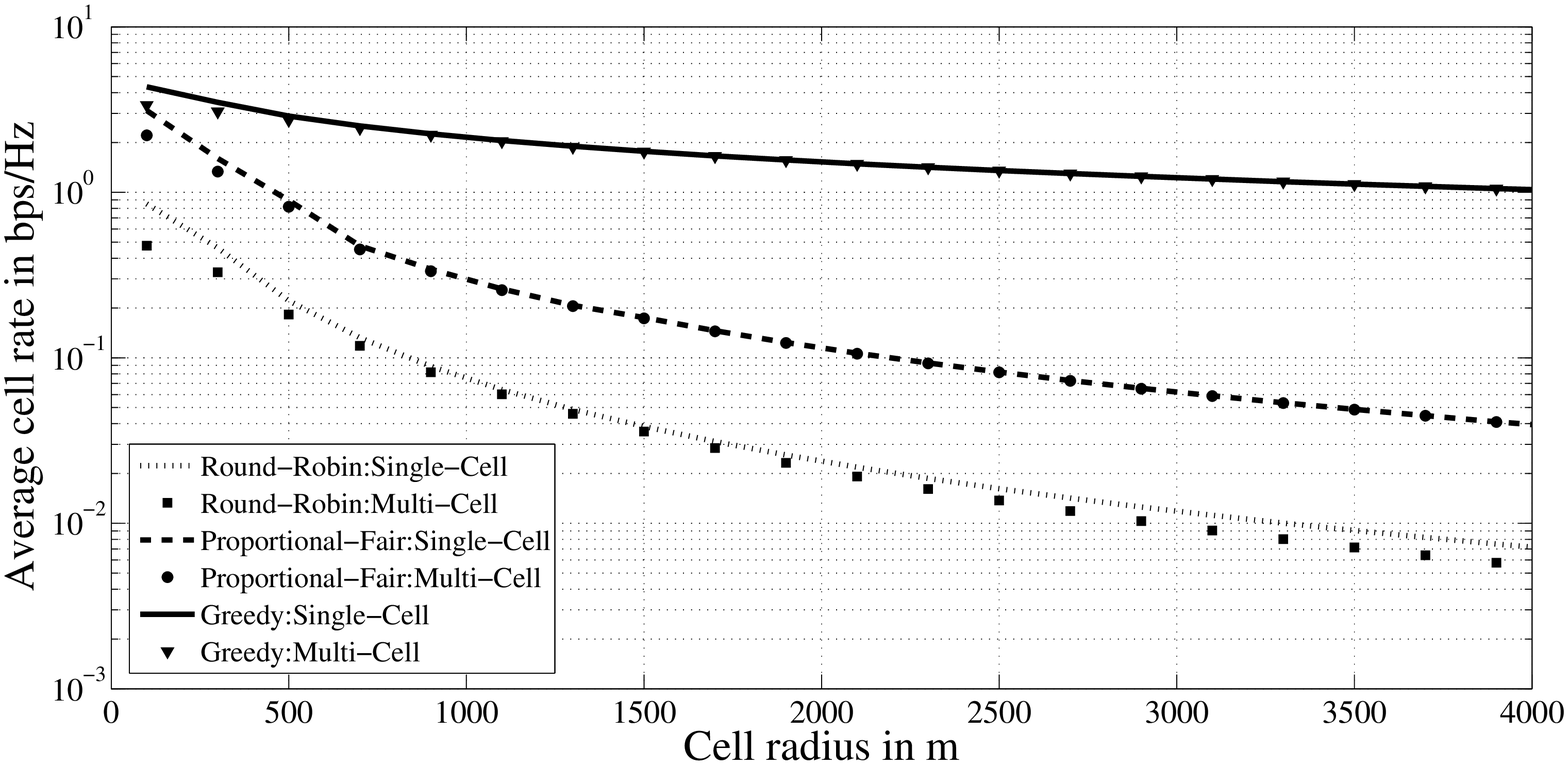}
    \caption{\small Capacity-coverage tradeoff results for various scheduling schemes with fixed BS powers independent of the cell radius.
      \label{figure5}}
  \end{center}\vspace{-0.5cm}
\end{figure}

In this section, we quantify the interplay between capacity and coverage (modeled via average cell rate and cell radius respectively) using the derived expressions. Moreover, we quantify the effect of ICI on the average cell rates as a function of the cell radius and scheduling scheme. We present results for all the previously discussed scheduling schemes; results can be easily generated for other scheduling schemes by varying the $\sigma$ parameter in~(\ref{eq:gen}). Fig.~\ref{figure5} quantifies the decay in average cell rate as the cell radius increases while maintaining a fixed power at the BS for both single and multi-cell scenarios. Since it is typical to increase the BS transmission power when the cell radius increases and vice versa, the fixed power assumption could be considered as per subcarrier. 

It can be seen that the impact of ICI is rather limited for proportional-fair and greedy schemes. However, this is not the case for round-robin scheduling where the gap gets bigger as the cell radius increases; this is because the relatively large number of scheduled cell-edge users will be subjected to lower SINR values. Note that the unit of the average cell rate is bps/Hz and, thus, the difference between single and multi-cell for round-robin scheduling is significant for high cell radii when multiplied by the utilized bandwidth.

It can be shown that the percentage loss in average rate for the proportional-fair scheme is maximal at lower cell radii ranging from approximately $30\%$ to negligible for radii greater than $1500 \,  \text{m}$.  The results of Fig.~\ref{figure5} clearly show the potential of using cells with small radii in terms of maximizing the cell rate for various scheduling schemes; this is inline with the current research and 3GPP standardization efforts towards small cell deployments. At the same time, the high percentage loss for small radii demonstrates that ICI has the highest relative negative impact on performance in small cell scenarios; this motivates the need for advanced coordinated scheduling and interference management techniques among neighboring small cells in 4G cellular networks.

\begin{figure}[t!]
  \begin{center}
    \includegraphics[width=3.4in]{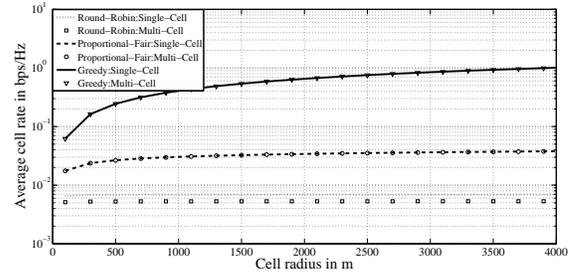}
    \caption{\small Capacity-coverage tradeoff results for various scheduling schemes with BS power adaptation as a function of the cell radius.
      \label{figure6}}
  \end{center}\vspace{-0.5cm}
\end{figure}

In Fig.~\ref{figure6}, we regenerate Fig.~\ref{figure5} assuming an adaptive power allocation scenario, where the power at the BS is scaled with the cell radius in such a way to maintain a constant pathloss for users at the cell boundary; this is motivated by the fact that smaller cells can be equipped with power amplifiers with lower power budgets. The reference power $P = 1$ W is assumed for a radius $\rho = 4000$~m. It is interesting to note the significant difference in performance compared to Fig.~\ref{figure5}. For the greedy scheme, the average cell rate increases as the cell radius increases due to the fact that we are still serving cell-center users however with higher powers. The proportional-fair scheme benefits from more user diversity since the number of users increases as the cell radius increases while preserving a fixed user density. However, the round-robin scheme has a relatively stable performance since users are selected arbitrarily where both signal power and interference power are scaled up with the cell radius. 

\section{Conclusion}
We presented analytical expressions for cell rate and intercell interference distributions in the downlink of cellular networks as a function of user scheduling schemes for general fading models. We discussed first a single cell scenario and validated the obtained analytical results using Monte-Carlo simulations. We extended the derivations to a multi-cell scenario and obtained semi-analytical expressions for the rate per given location inside the cell in addition to the overall average cell rate. Moreover, we presented a generic approach based on a truncated Gaussian distribution to mathematically model the user selection density function for a range of scheduling schemes. Closed form expressions were derived for a case study network scenario with a Rayleigh fading channel; several results were presented and discussed with focus on capacity-coverage tradeoffs in downlink cellular networks. 

\section{Acknowledgments}
This work was made possible by NPRP grant 4-353-2-130
from the Qatar National Research Fund (a member of The
Qatar Foundation). The statements made herein are solely the
responsibility of the authors.

%


\bibliographystyle{ieeetr}
\bibliography{references}

\end{document}